\documentclass[useAMS,usenatbib,fleqn]{mn2e}

\usepackage{graphicx}

\usepackage{mathptmx}

\DeclareMathAlphabet{\mathsc}{OT1}{cmr}{m}{sc}
\def\testbx{bx}%
\DeclareRobustCommand{\ion}[2]{%
\relax\ifmmode
\ifx\testbx\f@series
{\mathbf{#1\,\mathsc{#2}}}\else
{\mathrm{#1\,\mathsc{#2}}}\fi
\else\textup{#1\,{\mdseries\textsc{#2}}}%
\fi}

\newcommand{\ergs}{\mathrm{erg/s}}
\newcommand{\Msun}{\mathrm{M}_\odot}
\newcommand{\asec}{^{\prime\prime}}
\newcommand{\keV}{\mathrm{keV}}

\title[What obscures low X-ray-scattering AGN?]{What obscures low X-ray-scattering active galactic nuclei?\,\thanks{Based in part on data obtain with the Herschel observatory. Herschel is an ESA space observatory with science instruments provided by European-led Principal Investigator consortia and with important participation from NASA.}}

\author[S.~F. H\"onig et al.]{{\parbox{\textwidth}{S.~F. H\"onig$^{1}$, P. Gandhi$^{2}$, D. Asmus$^{3}$, R.~F. Mushotzky$^{4}$, R. Antonucci$^{5}$, Y. Ueda$^{6}$, K.~Ichikawa$^{6}$}}\vspace{0.4cm}\ \\
$^{1}$Dark Cosmology Centre, Niels Bohr Institute, Juliane Maries Vej 30, 2100 Copenhagen \O, Denmark; e-mail: shoenig@dark-cosmology.dk\\
$^{2}$Department of Physics, Durham University, South Road, Durham DH1 3LE, UK\\
$^{3}$Max-Planck-Institut f\"ur Radioastronomie, Auf dem H\"ugel 69, 53121 Bonn, Germany\\
$^{4}$University of Maryland, College Park, MD 20742, USA\\
$^{5}$University of California in Santa Barbara, Department of Physics, Broida Hall, Santa Barbara, CA 93106-9530, USA\\
$^{6}$Department of Astronomy, Kyoto University, Kyoto 606-8502, Japan\\
}

\date{Accepted 2013 November 14. Received 2013 November 14; in original from 2013 October 11}

\begin{document}

\pagerange{\pageref{firstpage}--\pageref{lastpage}} \pubyear{2013}

\maketitle

\label{firstpage}

\begin{abstract}
X-ray surveys have revealed a new class of active galactic nuclei (AGN) with a very low observed fraction of scattered soft X-rays, $f_\mathrm{scat}<$0.5\%. Based on X-ray modeling these ``X-ray new-type'', or low observed X-ray scattering (hereafter:``low-scattering'') sources have been interpreted as deeply-buried AGN with a high covering factor of gas. In this paper we address the questions whether the host galaxies of low-scattering AGN may contribute to the observed X-ray properties, and whether we can find any direct evidence for high covering factors from the infrared (IR) emission. We find that X-ray low-scattering AGN are preferentially hosted by highly-inclined galaxies or merger systems as compared to other Seyfert galaxies, increasing the likelihood that the line-of-sight toward the AGN intersects with high columns of host-galactic gas and dust. Moreover, while a detailed analysis of the IR emission of low-scattering AGN ESO 103--G35 remains inconclusive, we do not find any indication of systematically higher dust covering factors in a sample of low-scattering AGN based on their IR emission. For ESO 103--G35, we constrained the temperature, mass and location of the IR emitting dust which is consistent with expectations for the dusty torus. However, a deep silicate absorption feature probably from much cooler dust suggests an additional screen absorber on larger scales within the host galaxy. Taking these findings together, we propose that the low $f_\mathrm{scat}$ observed in low-scattering AGN is not necessarily the result of circumnuclear dust but could originate from interference of host-galactic gas with a column density of the order of $10^{22}\,\mathrm{cm}^{-2}$ with the line-of-sight. We discuss implications of this hypothesis for X-ray models, high-ionization emission lines, and observed star-formation activity in these objects.
\end{abstract}

\begin{keywords}
galaxies: active -- galaxies: nuclei -- galaxies: individual: ESO103--G35 -- galaxies: Seyfert -- X-rays: galaxies -- infrared: galaxies.
\end{keywords}

\section{Introduction}

\setcounter{footnote}{0}

The increasing sensitivity of X-ray observations enables us to study growing samples of active galactic nuclei (AGN) and test for extreme regions of the parameter space. Using the Burst Alert Telescope (BAT) onboard the \textit{Swift} X-ray satellite, we currently get access to an ever-increasing flux-limited catalog of AGN in the waveband of $14-195\,$keV, pushing the limit on hydrogen column densities toward $N_\mathrm{H} \sim 10^{25}$\,cm$^{-2}$.

\citet{Ued07} report \textit{Suzaku} X-ray follow-up observations in the $2-10$\,keV band of two highly obscured AGN identified in the \textit{Swift} BAT hard X-ray survey. Spectral modeling over 1--50\,keV reveals that those two AGN (ESO\,005--G4 and ESO\,297--G018) have extremely low observed X-ray scattering fractions of the order of 0.2\%. The authors interpret these objects as a class of X-ray ``new-type'' AGN where the nucleus is deeply embedded within a dusty torus with very small opening angle.

\begin{table*}
  \centering
  \caption{Thirteen X-ray low-scattering AGN ($f_\mathrm{scat}<0.5\%$) included in the BAT 9-month catalog and four additional ones identified in the XMM catalog.\label{tab:sample}}
  \begin{tabular}{l c c c c c c c}
\hline
object (alternative name)    & $\log N_\mathrm{H}$ & $\tau_\mathrm{silicate}$ &  $\alpha_\mathrm{MIR}^\mathrm{dereddened}$  & $\log L_{14-195\,\mathrm{keV}}$ & $\log \nu L^\mathrm{nuc}_\nu(12\,\micron)$ & inclination & comment\\ 
                                            &       (cm$^{-2}$)              &      &     &    (erg/s)    &   (erg/s)    &   (deg)     \\ \hline
\multicolumn{8}{c}{low-scattering AGN in the BAT catalog} \\ \hline
NGC 235A                           & 23.45$\pm$0.01  &$\ldots$&$\ldots$& 43.44   &  42.62    &  $90$  & merger  \\ 
MARK 348  (NGC 262)         & 23.20$\pm$0.02  &   0.20   & $-1.65$&  43.81   &  43.47    &   $68$   & flat-spectrum radio source \\
NGC 454                             & 23.20$\pm$0.01  &   0.22   & $-2.17$&  42.88   &  43.12    &   $90$ & pair/merger \\
ESO 297--G18                    & 23.62$\pm$0.04  &   0.08   & $-1.65$&  44.04   &  43.12    &   $90$ & edge-on \\ 
NGC 1142  (NGC 1144)       & 23.90$\pm$0.03  &$\ldots$&$\ldots$& 44.30   &$<$43.54&   $68$   & pair/merger \\ 
ESO 005--G4                      & 22.75$\pm$0.02  &$\ldots$&$\ldots$& 42.64   &$<$42.09&   $90$ & edge-on    \\ 
MARK 417                           & 23.92$\pm$0.05  &   0.05   & $-1.92$&  44.03   &   43.63    &   $38$ &    \\ 
ARK 347     (NGC 4074)       & 23.48$\pm$0.13  &$-$0.09~~~~& ~$-1.52^a$& 43.72   &   43.42    &   $64$ &    \\ 
ESO 506--G27                    & 23.89$\pm$0.04  &   0.93   & $-1.78$&  44.37   &   43.77    &   $84$ & edge-on   \\ 
NGC 4992                           & 23.84$\pm$0.03  &   0.94   & $-1.07$&  44.01   &   43.41    &   $61$ &    \\ 
ESO 103--G35                    & 23.33$\pm$0.05  &   0.73   & $-2.82$&  43.67   &   43.79    &   $84$ &    \\ 
NGC 7172                           & 22.91$\pm$0.19  &   2.59   & $-1.66$&  43.25   &   42.68    &   $53$ & edge-on, dust lane   \\ 
NGC 7319                           & 23.94$\pm$0.14  &$\ldots$&$\ldots$&  43.65   & $\ldots$ &   $53$ & interacting group  \\ 
\hline
\multicolumn{8}{c}{inclination for additional objects from \citet{Nog10} based on the XMM catalog } \\ \hline
 3C 33      & \multicolumn{5}{c}{} & $\ldots$  & point source  \\ 
 2MASX J02281350--0315023 & \multicolumn{5}{c}{} & $40$  & \\ 
 MCG+8--21--65               & \multicolumn{5}{c}{} & $87$  & edge-on \\ 
 IC 2461                              & \multicolumn{5}{c}{} & $90$  & edge-on \\ 
\hline
\multicolumn{8}{l}{Inclinations based on data from the \textit{HyperLeda} database \citep{Pat03} and estimated to be good to within about 5$^\circ$. They are calculated using} \\ 
\multicolumn{8}{l}{isophotal axis ratios and the morphological type of the galaxy. Luminosity are based on the latest Planck results: $H_0=67.3$\,(km/s)/Mpc,} \\
\multicolumn{8}{l}{$\Omega_m=0.315$, $\Omega_\Lambda = 0.685$. $^a$ not extinction-corrected because of negative $\tau_\mathrm{silicate}$.}
  \end{tabular}
\end{table*}

\citet{Ich12} use the more X-ray complete non-blazar BAT sample by \citet{Win09} and find that 13 out of the 128 AGN (=10\%) in the sample fall into the X-ray new-type category, defined as objects with soft X-ray scattering fraction $f^X_\mathrm{scat} < 0.5\%$ based on spectral modeling. The authors also compare the mid-IR fluxes obtained from the \textit{WISE}, \textit{AKARI}, and \textit{IRAS} all-sky surveys of these objects to ordinary type 1 and type 2 AGN and find a possible excess of $9\,\micron$ emission. This is tentatively interpreted as due to contaminating PAH emission from the $7.7\,\micron$ and $8.6\,\micron$ features, possibly indicating strong starburst activity in the hosts of these galaxies. 

ESO\,005--G4 is the infrared and X-ray brightest object of this class and may serve as a prototype. An optical spectrum in the 4000--5500\,$\AA$ range does not show any [\ion{O}{iii}] or H$\beta$ emission lines or an AGN continuum and is reminiscent of a non-AGN host galaxy with a spectral break at about 4500\,$\AA$ towards shorter wavelengths \citep{Ued07}. In fact, the classification of this galaxy as an AGN was only based on the X-ray detection in the \textit{Swift} BAT catalog, though later confirmed by high-ionization lines revealed in the mid-infrared \citep{Wea10}. A lower limit for the X-ray-to-[\ion{O}{iii}] luminosity ratio is given as $L(\mathrm{2-10\,keV})/L([\ion{O}{iii}](5007\,\mathrm{\AA})) > 2800$. Such a ratio seems to be common in the low-$f_\mathrm{scat}$ objects and is interpreted as evidence for a small torus opening angle within which the narrow-line region (NLR) can be hosted \citep{Nog10}.

In this paper we aim at investigating the source of absorption and infrared emission of the dust that is responsible for obscuring the AGN in X-ray new-type objects. Since it is unclear if these objects really consistute a new class of AGN, we will refer to them as ``low-scattering'' objects throughout the paper, which adequately relates to their observed characteristics. We particularly focus on the optical and infrared (IR) perspective. In Sect.~\ref{sec:sample} we introduce the X-ray complete sample drawn from the Swift/BAT 9-month catalog and present data on the optical and infrared emission. In Sect.~\ref{sec:discu} we analyse optical images of the host galaxies of low-scattering AGN, discuss their infrared properties compared to the complete BAT catalog, and determine physical parameters of the dust responsible for the IR emission of one of these objects (ESO~103--G35). We present an alternative explanation for the peculiar properties of low-scattering sources in Sect.~\ref{sec:newmod} and discuss their place in the broad context of AGN unification. Finally in Sect.~\ref{sec:summary}, we summarize our findings.

\section{Sample of X-ray low-scattering AGN}\label{sec:sample}

\subsection{X-ray sample selection and data}

Low-scattering AGN were identified and defined as objects with a soft X-ray scattering fraction of $f^X_\mathrm{scat} < 0.5\%$ \citep{Ued07}. No other criteria have been used to associate objects with this group initially. In order to obtain a meaningful sample and analyze the group statistically within the framework of AGN unification, it is necessary to select objects from an X-ray-complete catalog. Such a catalog is available from the Swift/BAT hard-X-ray mission that lists all objects detected in the 14--195\,keV band over a given period of time. Therefore, we follow \citet{Ich12} and use the 128 detections over the first 9 months of the Swift/BAT survey \citep{Tue08} that were associated with non-blazar AGN by \citet{Win09}. Although there are deeper samples available by now (most notably 22, 54, and 70 month catalogs), the convenience of this well-defined sample originates from the wealth of complementary data that have been collected for all objects \citep[e.g.][]{Win10,Ich12,Asm13a,Asm13b}.

\begin{figure*}
\begin{center}
\includegraphics[width=0.95\textwidth]{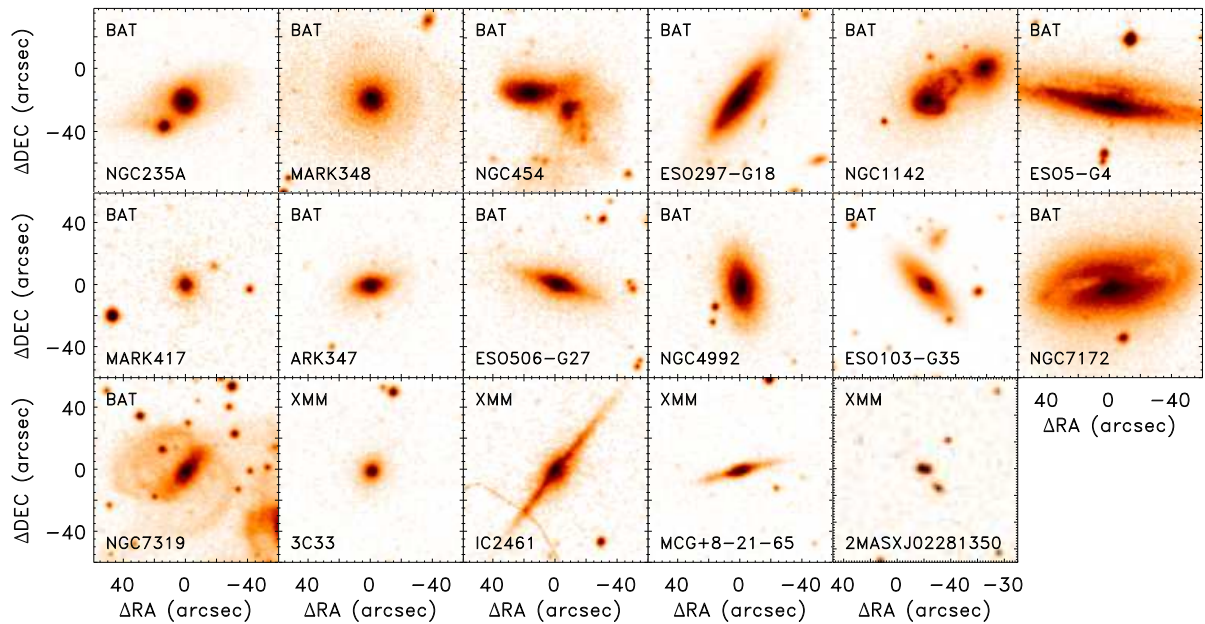}
\end{center}
\caption{DSS red images from the second DSS survey of the thirteen low-scattering AGN from the 9-month \textit{Swift/BAT} catalog and 4 objects from the \textit{XMM} source catalog (DSS-1 survey image for 2MASX J02281350--0315023).}\label{fig:images}
\end{figure*}

13 objects have been identified as low-scattering AGN out of the 128 non-blazar AGN in the Swift/BAT 9-month catalog \citep{Ich12}. We list all these objects in Table~\ref{tab:sample}. In addition we computed hard-X-ray luminosities, $\log L_{14-195\,\mathrm{keV}}$, using fluxes reported in the 22-month Swift/BAT catalog by \citet{Tue10}, which have generally higher signal-to-noise (S/N) than in the 9-month catalog due to the longer integration times. The only exception is NGC 454 which was detected at S/N=4.54 in the 9-month sample but does not show up above the 22-month sample limit of S/N=4.8. Table~\ref{tab:sample} also includes X-ray column densities obtained from spectral fitting by \citet{Win09}.

In Table~\ref{tab:sample} we also list four more low-scattering objects that have been identified by \citet{Nog10} in the \textit{XMM} AGN source catalog \citep{Wat09}. This catalog, however, is not complete since it was extracted from all pointings of \textit{XMM} with different exposure times on selected areas of the sky. Therefore we do not include these objects in our analysis related to unification issues/AGN statistics, but will use them when dealing with specific properties of low-scattering AGN as an object class. In particular, we will use these objects in Sect.~\ref{sec:stat} to increase number statistics when studying the host galaxies of low-scattering AGN.

\subsection{Complementary optical and infrared data}\label{sec:optir}

In this paper we want to address the questions if we can find evidence of a large dust covering factor of the low-scattering AGN and/or if other factors may also play a role in the observed low soft-X-ray scattering fractions. Here, we primarily focus on the optical and infrared regimes to find clues on the nature of the obscuration associated with these objects and for direct evidence of large covering factors by IR reemission that is significantly stronger than in typical Seyfert galaxies compared to the AGN luminosity traced by the BAT $14-195$\,keV emission (see Sect.~\ref{sec:mirx} for details).

The IR emission of the low-scattering objects has been previously discussed by \citet{Ich12}. They analyzed broad-band \textit{2MASS} photometry in the $J$, $H$, $K$ bands as well as 9\,$\micron$, 18\,$\micron$, and 90\,$\micron$ photometry obtained by the \textit{AKARI} satellite mission and found possible evidence for enhanced star-formation as compared to other type 1 and type 2 AGN. Given the low spatial resolution and limited spectral coverage of these data, it is, however, difficult to disentangle galaxy-scale emission from the circumnuclear environment. Therefore we chose another strategy: we searched the \textit{Spitzer} archive for IRS spectra and found suitable low-resolution data for 12 of the 13 sources. Although at this spatial resolution host and AGN emission are still confused, the spectral coverage allows us to use a decomposition technique to isolate the nuclear IR emission. For that, we computed a star-formation template by comparing objects that show star-formation signatures in IRS spectra but are devoid of any such features in our ground-based VLT/VISIR spectra with $\sim$10 times better spatial resolution \citep{Hon10a}. The template covers the $8-13\,\micron$ wavelength region. We determined the 11.3\,$\micron$ PAH equivalent width in all the IRS spectra and, accordingly, scaled and subtracted the template. The resulting spectra resemble the ground-based high-angular resolution spectra of AGN \citep[i.e. they are free of any star-formation features; e.g.][]{Pac05,Mas06,Mas09,Hon10a} and we consider them representative for the AGN IR emission. 

From these spectra we computed three quantities that are listed in Table~\ref{tab:sample}: (1) the nuclear 12\,$\micron$ luminosities $\nu L^\mathrm{nuc}_\nu(12\,\micron)$ of the AGN; (2) the optical depth $\tau_\mathrm{silicate}$ of the silicate feature located at about $10\,\micron$; and (3) the spectral slope $\alpha$ in the mid-IR. The latter two quantities were obtained by applying the same method used for ground-based spectra in \citet{Hon10a} with the same wavelength coverage. Where the star-formation templates accounted for all emission in the $8-13\,\micron$ band, we provide upper limits to the mid-IR luminosity.

Finally, in order to test the origin of the obscuration, we inspect optical DSS images of the complete sample to look for any host galaxy feature that may intersect with the nucleus. The images of all thirteen BAT sources and four additional \textit{XMM} sources are shown in Fig.~\ref{fig:images}. We selected images from the rather old DSS survey since they contain the whole sample uniformly, while newer surveys, such as SDSS, miss several ones.

\section{Discussion}\label{sec:discu}

\subsection{Statistics of host galaxy inclination}\label{sec:stat}

One question that has not been addressed for low-scattering AGN yet is whether their peculiar properties are related to the host galaxies. Indeed, \citet{Win09} note that while AGN with low X-ray column densities are preferentially found in host galaxies with low inclination, those at higher column densities are found in galaxies of any inclination. This suggests that the host galaxy can contribute at least in part to obscuration in AGN. A recent comparison of 10\,$\micron$ silicate absorption features, X-ray column densities, and host inclinations confirms this idea for highly-obscured objects \citep{Gou12}. The authors showed that deep silicate absorption typically appears in AGN where the galaxy is edge-on or engaged in a merger. In such galaxies, it is very likely that dust lanes are projected onto the nuclear region. Finally, using an isotropically selected sample, \citet{Kin00} showed that type 2 AGN do not reside in galaxies with a preferred inclination. In their sample, there seems to be a lack of type 1 AGN in edge-on galaxies. However, this can be explained by obscuration preferentially taking place on galactic scales that leads to an ``intrinsic type 1'' being classified as a type 2. When taking this effect into account, Kinney et al. conclude that the orientation of the AGN and the host galaxy are completely random and not correlated. This is consistent with \citet{Kee80} who also noted a substantial deficiency of Seyfert 1 galaxies with nearly edge-on host inclinations \citep[see also][]{Law82,Kir90}.  From these studies we conclude that while there is no correlation between host and AGN orientation, an inclined host can influence the classification of an AGN, making it appear as a type 2 independent of its actual intrinsic/torus orientation. We take this as a motivation to quantitatively assess if host galaxy inclination might play a role in the low-scattering phenomenon. 

In Table~\ref{tab:sample} we list the inclinations of the 13 low-scattering AGN host galaxies in the BAT sample as well as for 3 out of the 4 XMM objects. These inclinations have been taken from the HyperLeda catalog and are calculated based on optically-decomposed galaxy morphologies, i.e. major- and minor axis as well as Hubble type (see the catalog's website for further details). No inclination was listed in the database for 3C~33, which only appears as a point source in the DSS image. The inclinations from the database match the overall impression of the DSS images in Fig.~\ref{fig:images} well. However, in cases of merger systems, the host galaxy can be severely distorted and, therefore, the Hubble type may not be uniquely determined. 

The question we want to answer is whether the host galaxies of low-scattering objects show any preferred orientation. The low-scattering sample for this type of investigation does not require to be selected from a complete AGN parent sample, as long as the parent sample does not favor any host inclination. Neither the BAT nor the XMM catalog bias towards any host galaxy orientation in the first place. Hence, we combine the BAT and XMM objects in this study to increase sample statistics to a total of 16 objects with known inclinations. When considering the results of \citet{Kin00}, \citet{Win09}, and \citet{Gou12}, however, it may be possible that some AGN in edge-on galaxies are missing from the samples because of higher obscuration and, therefore, lower X-ray fluxes. In this respect, our sample might be slightly biased towards lower host inclination, which we will discuss later.

Looking at the DSS images in Fig.~\ref{fig:images}, we find it remarkable that at least 12/16 low-scattering AGN may be clearly classified as edge-on, merging, or interacting by visual inspection of the optical images only. More quantitatively, in a random inclination distribution $g(\phi\ge 80^\circ) = 17\%$ of objects are expected to be seen at inclinations $i\ge80^\circ$. Table~\ref{tab:sample} shows that the host galaxies of the low-scattering sample display such high inclinations at a rate of 50\% (8/16). Conversely, it suggests a deficiency of low-inclination objects among the low-scattering class.

\begin{figure}
\begin{center}
\includegraphics[width=0.5\textwidth]{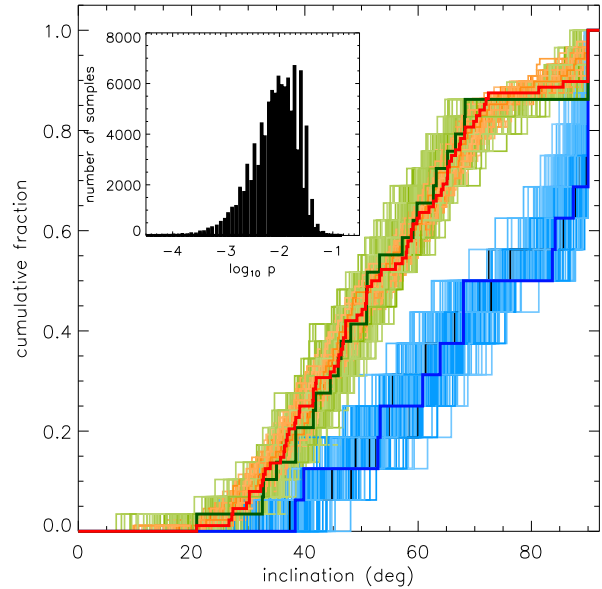}
\end{center}
\caption{KS test for the host-galaxy inclination distribution of low-scattering AGN. The dark-red line shows the observed cumulative inclination distribution of the BAT 9-month AGN without low-scattering AGN while the dark-blue line is the cumulative inclination distribution for the low-scattering sources. The green line represents all BAT 9-month AGN in the same column density range as the low-scattering AGN, $22.5 \le \log N_\mathrm{H}(\mathrm{cm}^{-2} < 24.0$. For all three sub-samples we simulated 100\,000 random realizations assuming an inclination error of 5$^\circ$ (1$\sigma$). These realizations are shown as light-red, light-blue, and light-green for the three sub-samples, respectively.}\label{fig:ks}
\end{figure}

In order to be more quantitative, we want to test if the low-scattering AGN prefer any host inclinations. In principle, host galaxy inclinations should follow a random distribution where the fraction of objects $g(\phi\ge i)$ seen at an inclination $i$ or greater is given by the solid angle covered by these angles as $g(\phi\ge i) = \cos i$. However, since we do not know how well this distribution is reproduced by the method used to determine inclinations in the HyperLeda database, we took a different approach. We extracted inclinations of the whole BAT 9-month catalog from the HyperLeda catalog and divided the objects into two sub-samples: (1) the thirteen low-scattering AGN; (2) all other 115 AGN. Then we can compare the inclination distributions of both samples by a two-sided KS test. If the low-scattering AGN do not favor any specific host inclination, we expect that both samples follow the same distribution. Although no errors are provided by HyperLeda, we also consider a 5$^\circ$ uncertainty (1-$\sigma$ error; Gaussian error distribution) on each inclination value and simulate 100\,000 realizations for the two sub-samples. The results are shown in Fig.~\ref{fig:ks}. Based on the KS test, we find a median probability that two random draws from a single parent distribution would produce the observed distributions of low-scattering AGN and all BAT sources without low-scattering AGN of only $p=9.7 \times 10^{-3}$ (significance 2.6$\sigma$). Therefore, we conclude that the low-scattering AGN favor certain inclinations. Indeed, Fig.~\ref{fig:ks} clearly shows that the low-scattering AGN class populates the high-inclinations that are missing in the sub-sample where the low-scattering objects have been excluded. 

One selection effect not considered in this comparison is the fact that low-scattering AGN can only be identified by spectral modeling if absorption is detected. This favors ``intrinsic'' type 2 AGN against type 1 AGN. We address this bias by also comparing the low-scattering AGN to all BAT AGN in the same column density range as the low-scattering AGN, $22.5 \le \log N_\mathrm{H}(\mathrm{cm}^{-2}) < 24.0$, but excluding the low-scattering objects from the sample. The cumulative distribution of this obscured AGN comparison sample is also shown in Fig.~\ref{fig:ks}. We find a slightly increased null hypothesis probability of $p=2.2 \times 10^{-2}$ when comparing this sample to the low-scattering AGN. However, this lower significance is only a result of the decreased sample statistics with respect to the full BAT 9-month catalog. Indeed, the KS statistics remains essentially unchanged ($d=0.42$ for the full sample vs. $d=0.44$ for the column-restricted sample). Therefore we conclude that the preference of high host inclination of low-scattering AGN is present even when comparing them with objects in the same column density range.

In summary, we find evidence that the low-scattering AGN are preferentially hosted in edge-on galaxies. This implies that absorption within the host galaxy, not intrinsic to the immediate AGN environment, might play an important role in the observed properties of this class of objects. As mentioned before, it is well possible that the X-ray samples slightly bias against AGN in galaxies at high inclination and, thus, potentially against low-scattering objects. Therefore, the evidence for high-inclination preference may become stronger as more such objects are identified with BAT.

\subsection{Determining the IR covering factor in ESO\,103--G35}\label{sec:eso103}

One way to directly test the hypothesis of a deeply-buried AGN in low-scattering objects is determining the dust covering factors. For that we need an IR SED over a wide wavelength range that is dominated by dust reemission of the big-blue bump of the hidden AGN. In practice, high-angular resolution data are required to minimize host-galactic contamination. 

\begin{table}
  \centering
  \caption{Broadband IR photometry of ESO~103--G35.\label{tab:irphot}}
  \begin{tabular}{l c c c}
\hline
band            & obs. wavelength$^a$     & flux                   & Ref. \\
                     & ($\micron$)          & (mJy)                  &              \\ \hline
2MASS-$J^b$                 & 1.24$\pm$0.16 &      3.2$\pm$0.7   & (1) \\
2MASS-$H^b$                & 1.66$\pm$0.25 &      2.6$\pm$1.0  & (1) \\
2MASS-$K^b$                &  2.16$\pm$0.26   &      3.8$\pm$0.8   & (1) \\
WISE-W1                     & 3.4$\pm$0.3        &     41.6$\pm$0.9   & (2) \\
WISE-W2                     & 4.6$\pm$0.5        &     96.4$\pm$1.9   & (2) \\
VISIR PAH1                  & 8.59$\pm$0.42 &   258$\pm$24        & (3) \\
AKARI-S09                  & 9.4$\pm$2.1        &   300$\pm$25       & (4)\\
VISIR SIV                      & 10.49$\pm$0.16 &   243$\pm$15        & (5) \\
VISIR PAH2                  & 11.25$\pm$0.59 &   405$\pm$19        & (3) \\
IRAS 12\,$\micron$    & 11.75$\pm$3.3      &   612$\pm$43       & (6) \\
WISE-W3                     & 11.6$\pm$2.7      &   522$\pm$7          & (2) \\
VISIR Ne2ref2               & 13.04$\pm$0.22 &   622$\pm$49   & (5) \\
AKARI-S18                  & 19.6$\pm$5.6      & 1446$\pm$12       & (4) \\
WISE-W4                     & 22.1$\pm$2.0      & 2069$\pm$6         & (2) \\
IRAS 25\,$\micron$     & 25$\pm$6           & 2360$\pm$120     & (6) \\
MIPS 24\,$\micron$     &  27$\pm$5          & 1693$\pm$81 & (7) \\         
IRAS 60\,$\micron$     & 60$\pm$20         & 2310$\pm$120     & (6) \\
AKARI-N60                   & 66$\pm$9           & 1833$\pm$51       & (4) \\
MIPS 70\,$\micron$     &  70$\pm$10        & 1774$\pm$112 & (7) \\  
PACS 70\,$\micron$    & 70$\pm$12         & 1732$\pm$88       & (8) \\
PACS Spectrum             &  73.5$\pm$0.5    & 1433$\pm$176 & (9) \\
PACS Spectrum             &  85.5$\pm$0.5    & 1294$\pm$100 & (9) \\
PACS Spectrum             &  91.2$\pm$0.25    & 988$\pm$176 & (9) \\
AKARI-S90                   & 94$\pm$13         & 1127$\pm$77       & (4) \\
PACS Spectrum             &  94.0$\pm$0.25    &   989$\pm$324 & (9) \\
IRAS 100\,$\micron$   & 102$\pm$19         & 1050$\pm$260     & (6) \\
PACS Spectrum             &  115.0$\pm$0.5  &   832$\pm$98 & (9) \\
AKARI-L140                  & 145$\pm$30         & 1828$\pm$906       & (4) \\
PACS Spectrum             &  146.8$\pm$0.5  &   612$\pm$97 & (9) \\
PACS 160\,$\micron$  & 160$\pm$40       &   551$\pm$34       & (8) \\
PACS Spectrum             &  171.1$\pm$0.5  &   470$\pm$63 & (9) \\
PACS Spectrum             &  178.5$\pm$0.5  &   539$\pm$120 & (9) \\
PACS Spectrum             &  182.5 $\pm$0.5  &   406$\pm$235 & (9) \\
SPIRE 250\,$\micron$  & 250$\pm$40       &   157$\pm$10     & (10) \\
SPIRE 350\,$\micron$  & 350$\pm$50       &     66$\pm$8 & (10) \\
SPIRE 500\,$\micron$  & 500$\pm$100     &$      <$88 & (10) \\
\hline
\multicolumn{4}{l}{\textbf{References} --- (1) \citet{Skr06}, (2) WISE catalog,} \\
\multicolumn{4}{l}{(3) \citet{Asm13a}; (4) AKARI catalog, (5) \citet{Gan09},} \\
\multicolumn{4}{l}{(6) IRAS point source catalog, (7) \citet{Tem09}, (8) Melendez } \\
\multicolumn{4}{l}{et al., in prep., (9) this work; (10) Shimizu et al., in prep.;} \\
\multicolumn{4}{l}{\textbf{Notes} --- $^a$ central wavelength and band width of the filter;} \\
\multicolumn{4}{l}{$^b$ galaxy-subtracted nuclear flux}
  \end{tabular}
\end{table}

\begin{figure}
\begin{center}
\includegraphics[width=0.5\textwidth]{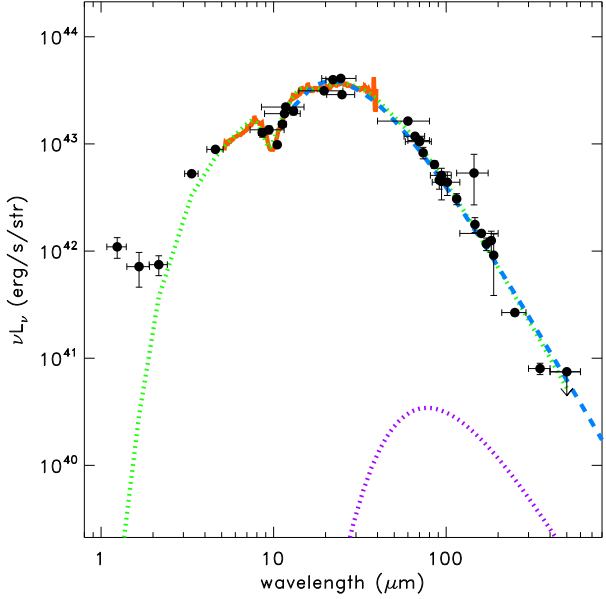}
\end{center}
\caption{IR SED of ESO\,103--G35 from 1\,$\micron$ to 500\,$\micron$. Aside from photometry from different telescopes and instruments (black circles with error bars; see Table~\ref{tab:irphot} for details), a \textit{Spitzer} IRS spectrum is shown as a red-solid line. The green-dotted line represents the model SED that is used for integration of the total IR emission. The blue-dashed line is a gray-body fit to the peak and Rayleigh-Jeans tail of the SED. We also show the expected emission spectrum from a cool host-galactic dust cloud projected onto the nucleus (violet-dotted line; see text for details).}\label{fig:eso103}
\end{figure}

Out of our low-scattering sample, we have the best collection of these kind of data available for ESO\,103--G35. Indeed, we acquired new Herschel photometry of this source that helped to significantly improve the wavelength coverage of the IR SED. In Fig.~\ref{fig:eso103} we present the nuclear IR SED of this source in the range of 1\,$\micron$ to 500\,$\,\micron$. The far-IR part of the data were observed with the ESA Herschel Space Observatory \citep{Pil10}, in particular the PACS \citep{Pog10} and SPIRE \citep{Gri10} instruments. Aside from the photometry listed in Table~\ref{tab:irphot}, we also overplot a \textit{Spitzer} IRS spectrum. The SED shows a strong rise from the near-IR to mid-IR, which is typical for type 2 AGN. Longward of about 20\,$\micron$, the SED turns over and decreases toward the far-IR/sub-mm regime. This turn-over in the mid-IR without and secondary bump in the far-IR is very indicative that most or all of the IR emission is, indeed, originating from the AGN; star formation would peak at far-IR wavelengths. Further evidence for the AGN domination comes from the fact that both high-angular and low-angular resolution photometry ($\sim0\farcs4 - 40\asec$) connect smoothly to a consistent and continuous SED. In addition, a fit of our star-formation template used in Sect.~\ref{sec:optir} reveals a limit on star-formation of $\la$6\%.

A silicate absorption feature is seen at $\sim10\,\micron$ that could be either coming from self-absorption within the IR emitting dust or an external cold absorber. Given the relatively deep feature and the high galaxy inclination, we suspect that most of the absorption responsible for the strength of the feature is not related to the mid-IR emitting dust. Moreover the near-IR photometry indicates an upturn toward shorter wavelengths. We are, however, not sure if this is related to nuclear emission \citep[e.g. the big-blue bump may cause such a feature; e.g.][]{Kis08,Kis09b} or if it is rather related to stellar emission that remains after host galaxy subtraction. In either case, since the near-IR emission flux is lower by about 1.5 orders of magnitude compared to the peak of the IR emission in the mid-IR, it will not significantly affect the bolometric IR luminosity we will extract from this SED.

In order to constrain the covering factor, we have to integrate the whole IR emission. For that we model the IR emission with a simple temperature gradient model that captures the essence of more complex clumpy torus models \citep[see][for a discussion]{Hon10b}. For our purpose, the details of the model do not matter since the only purpose is to reproduce the IR SED continuously without any intention to interpret the parameters. We note, however, that we have to add an extinction source to reproduce the silicate feature. The fit to the SED is shown in Fig.~\ref{fig:eso103} as a green-dotted line. Integration of the continuous model SED results in a bolometric IR luminosity of $L_\mathrm{IR} = (2.04\pm0.15) \times 10^{44}\,\ergs$.

We also need a way to determine the luminosity of the big-blue bump, $L_\mathrm{BBB}$, to constrain the covering factor. Since the AGN is heavily obscured, it is not possible to get this information directly from the optical/UV SED, which is dominated by starlight. However, the nucleus is obscured by a Compton-thin gas column, meaning that we can use the X-ray luminosity as a proxy to estimate $L_\mathrm{BBB}$. In \citet{Gan09}, we refer to an absorption-corrected X-ray luminosity $\log L_\mathrm{2-10\,\keV}(\ergs) = 43.4 \pm 0.2$ (corrected for the different cosmological parameters). Using the relation between $L_\mathrm{2-10\,\keV}$ and $L_\mathrm{BBB}$ by \citet{Mar12}, we obtain $\log L_\mathrm{BBB}(\ergs) = 44.6 \pm 0.3$.

Based on our values for $L_\mathrm{BBB}$ and $L_\mathrm{IR}$, we calculate the (bolometric) covering factor $C_\mathrm{bol} = L_\mathrm{IR}/L_\mathrm{BBB} = 0.5^{+0.5}_{-0.3}$. While this is the most direct way of testing the buried AGN scenario, the constraint we get is rather weak given the large uncertainty on $L_\mathrm{BBB}$. Therefore we cannot use this method in the case of ESO 103--G35 as an individual source. However, it may become more relevant in the future when good IR SEDs of the other low-scattering AGN become available and a sample average can be determined with smaller errors. 

Given this prospective, we want to remind the reader of a fundamental issue of covering factors. The reason why we call the covering factor determined this way as ``bolometric'' is related to the anisotropic radiation from the AGN. In non-spherically obscured AGN, i.e. as in a toroidal geometry, some degree of anisotropy of the emerging IR can be expected depending on the line-of-sight toward the obscurer, e.g. as seen in radiative transfer models \citep[e.g.][]{Sch05,Hon06,Sch08,Hon10b} and observations \citep[e.g.][]{Buc06,Gan09,Hon11,Asm11}. Therefore, the observationally-determined covering factor depends on viewing angle and potentially details of the dust distribution \citep[see also][]{Eli12}. Additionally, the actual distribution of optically thick and thin dust may be more complex, making the definition of a covering factor even more ambiguous \citep[see][]{Hon12,Hon13,Bur13,Kis13,Tri13}.

Having a well-covered IR SED also allows us to constrain the dust mass that participates in IR emission. We fitted a gray body spectrum to the peak region and Rayleigh-Jeans tail of the IR SED of ESO~103--G35 by fixing the dust opacity at 100\,$\micron$ to $\kappa_\mathrm{100} = 0.4\,\mathrm{m^2/kg}$ and leaving the spectral index of the wavelength-dependence of $\kappa_\lambda \propto \lambda^\alpha$ and the peak emission temperature $T_\mathrm{peak}$ as free parameters. As a result, we obtained a peak temperature $T_\mathrm{peak} = 169\pm2$\,K and $\alpha=0.0\pm0.1$, and accordingly a dust mass of $M_\mathrm{dust} = (1.4\pm0.4) \times 10^4\,\Msun$. We can estimate the scale at which the bulk of this mass is located, under the assumption that the IR-emitting dusty gas is identical to the X-ray obscuring material, as 
\begin{equation}
r \sim 17\,\mathrm{pc} \cdot (M_4/(N_{\mathrm{H},23}\,\xi_{0.01}))^{1/2}
\end{equation}
for a dust mass $M_4 = M_\mathrm{dust}/10^4\,\Msun$, a hydrogen column density $N_{\mathrm{H},23} = N_\mathrm{H}/10^{23}\,\mathrm{cm}^{-2}$, and a dust-to-gas ratio $\xi_{0.01} = \xi/0.01$. For ESO~103--G35 we obtain $r \sim 13\,\mathrm{pc}$, meaning that the bulk of the IR-emitting dust mass is located on scales commonly associated with the torus.

We want to emphasize, however, that this does not rule out a cool host-galactic screen absorber that is responsible for the strong silicate absorption feature, optical extinction, and potentially affecting the soft X-ray spectrum, since a column density of the order $10^{22}\,\mathrm{cm}^{-2}$ in a galactic dust cloud is sufficient to explain these. In Fig.~\ref{fig:eso103} we also show the emission from a host-galactic dust cloud with a temperature of 50\,K, the same size as the nuclear dust (to allow complete covering) and a tenth of the column density. As evident, the contribution to the emission is $\ll$1\% at any wavelength in the range we show. This illustrates that a significant amount of host-galactic cool dust can still be projected on the nucleus without showing up as emission in the 1$-$500\,$\micron$ SED

To summarize, we conclude that the IR emission in ESO~103--G35 does not allow for a strong constraint on the covering factor of dust. We conclude, however, that the IR \textit{emitting} dust is located within scales of $\la$10\,pc, consistent with the classical torus and IR interferometry observations that resolve these spatial scales \citep[e.g.][]{Kis11b,Bur13}. The observations do not rule out a cold screen absorber that could be responsible for the lack of soft X-rays and strong optical extinction. In fact, a deep silicate absorption feature in the IR SED lends support to this idea since dust on sub-parsec scales would probably be warm enough ($>$100\,K) to contribute to silicate emission and fill up (part of) the absorption feature \citep[see also][]{Gou12}.

\subsection{Low-scattering AGN an the mid-IR/X-ray correlation}\label{sec:mirx}

Since we do not have the same quality IR SED for the other objects as we have for ESO~103--G35, we are not able to directly determine the dust covering factors of the whole sample. However, if the covering factors in these objects are really high, we would expect that the ratio of a dust reemission tracer to an AGN luminosity tracer is particularly high compared to a more general AGN sample. The ratio of high-angular resolution mid-IR emission to (absorption-corrected) X-ray emission may serve as such a covering factor tracer ratio. Over the last years a baseline for this ratio has been established in the form of a tight correlation between the mid-IR and X-ray emission from scales $<$100\,pc around the AGN \citep{Hor06,Hor08,Gan09,Asm11}. In \citet{Asm13b}, this ratio is investigated for the whole BAT 9-month sample and, unlike previous publications, considers both the $2-10$\,keV as well as the $14-195$\,keV energy regions. Moreover, the ratio is also studied as a function of Hydrogen column density $N_\mathrm{H}$.

\begin{figure}
\begin{center}
\includegraphics[width=0.5\textwidth]{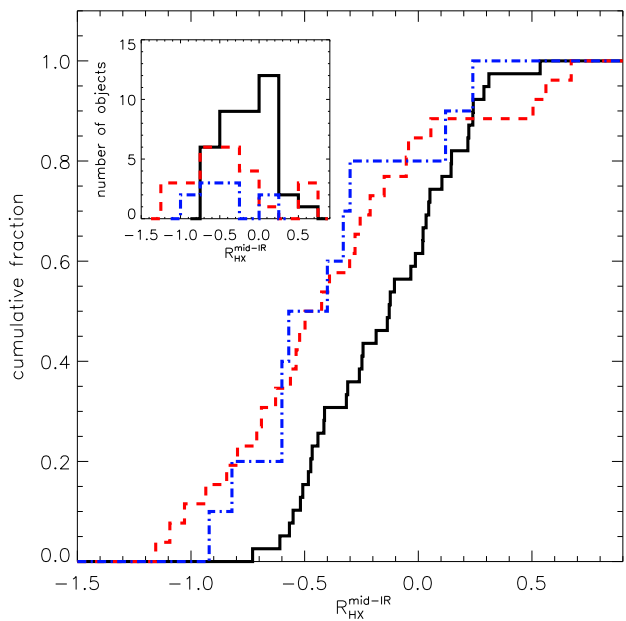}
\end{center}
\caption{KS test for the distribution of mid-IR/hard X-ray ratio $R^\mathrm{mid-IR}_\mathrm{HX}$ of BAT AGN. The black, solid line shows the cumulative distribution of all AGN in the BAT 9-month catalog with X-ray column densities $\log N_\mathrm{H}(\mathrm{cm}^{-2}) < 22.5$. The blue, dash-dotted line represents all low-scattering AGN identified in the BAT catalog, while the red, dashed line shows all BAT AGN with $22.5 \le \log N_\mathrm{H}(\mathrm{cm}^{-2})< 24.0$ excluding the low-scattering sources. In the small inset we plot the actual source number distributions of all three samples.}\label{fig:ks_ratio}
\end{figure}

For the present analysis we are mostly interested in a comparison of the mid-IR/hard X-ray ratio $R^\mathrm{mid-IR}_\mathrm{HX}$ of the complete BAT catalog --- mainly consisting of local Seyfert galaxies --- with the low-scattering AGN. If these were indeed deeply buried, we would expect that their mid-IR/X-ray ratio is larger than for the average Seyferts, since the reprocessing surface (mid-IR) is larger compared to other AGN with the same intrinsic luminosity (X-ray). We produce three different subsamples from the full BAT catalog data in \citet{Asm13b}: (a) all BAT sources with $\log N_\mathrm{H}(\mathrm{cm}^{-2}) < 22.5$ (mean luminosity $\log \left< L_{14-195\,\mathrm{keV}}(\mathrm{erg/s})\right> = 43.8\pm0.7$); (b) all BAT sources with $22.5 \le \log N_\mathrm{H}(\mathrm{cm}^{-2}) < 24.0$ that are not low-scattering AGN  ($\log \left<L_{14-195\,\mathrm{keV}}(\mathrm{erg/s})\right> = 43.5\pm1.0$); (c) all low-scattering AGN in the BAT sample ($\log \left<L_{14-195\,\mathrm{keV}}(\mathrm{erg/s})\right> = 43.7\pm0.5$). As shown in Table~\ref{tab:sample}, the low-scattering objects have Hydrogen column densities in the range of approximately $22.5 \le \log N_\mathrm{H}(\mathrm{cm}^{-2}) < 24.0$. 

To answer the question if the low-scattering AGN show any difference with respect to other Seyfert galaxies, we compare $R^\mathrm{mid-IR}_\mathrm{HX}$ of the low-scattering sample to all other BAT AGN within the same column density range. For that we perform a two-sided KS test on the distribution of $R^\mathrm{mid-IR}_\mathrm{HX}$. The cumulative distributions of the low-scattering sample (blue, dash-dotted) and the BAT AGN in the same $N_\mathrm{H}$ range (red, dotted) are shown in Fig.~\ref{fig:ks_ratio}. The KS test reveals a probability $p=0.95$ that random draws from a single parent distribution would be as strongly correlated with $R^\mathrm{mid-IR}_\mathrm{HX}$ as both observed samples at hand. The significance may be improved once a larger sample is available (see inset for the number distribution). 

We use the BAT objects at lower $N_\mathrm{H}$ as a control sample (black, solid line in Fig.~\ref{fig:ks_ratio}). These objects are dominated by AGN classified optically and in the X-rays as type 1 AGN. Both samples with $\log N_\mathrm{H}(\mathrm{cm}^{-2}) \ge 22.5$ show a tendency for lower average $R^\mathrm{mid-IR}_\mathrm{HX}$ values \citep[see also][]{Asm13b} which is consistent with the idea of anisotropic absorption and/or IR emission, e.g. as caused by the AGN torus or by host-galactic dust lanes\footnote{We want point out that our interpretation of the differences in $R^\mathrm{mid-IR}_\mathrm{HX}$ between low- and high-$N_\mathrm{H}$ objects depends on the assumption that the 14$-$195\,keV radiation is emitted more isotropically than the 12\,$\micron$ radiation.}. The actual degree of anisotropy, its consequences on covering factors, and its relation to the nature of the dusty torus are long-debated issues and we refer the reader to other work \citep[e.g. see][]{Hor08,Gan09,Lev09,Hon11,Alo11,Asm11,Eli12,Asm13b}. The main point here is that the ratio of both the low-scattering and ``normal'' AGN at the same column density show about the same distribution.

Having in mind that $R^\mathrm{mid-IR}_\mathrm{HX}$ is a tracer ratio for the covering factor, we conclude that we do not find any indication that low-scattering objects show any different covering factors than other objects with the same X-ray obscuration. Therefore we conclude that there is most likely no difference in mid-IR/X-ray ratios between low-scattering and non-low-scattering AGN. 

\subsection{Absorption and the mid-IR spectral slope}

One of the signs that might have been taken as evidence for low-scattering AGN being deeply buried is the deep $10\,\micron$ silicate absorption feature seen in Spitzer spectra of some sources. A complication when determining the strength of these features arise from PAH emission that is frequently present in spectral data of lower angular distribution. Not taking this into account would lead to an overestimation of the silicate absorption arising from the active nucleus. As described in Sect.~\ref{sec:optir}, we determined the optical depth in the silicate feature $\tau_\mathrm{silicate}$ without PAH contamination for the 13 low-scattering AGN in the BAT 9-month catalog. The values listed in Table~\ref{tab:sample} cover a range of $-0.09 \le \tau_\mathrm{silicate} \le 2.59$, with most of the objects showing $\tau_\mathrm{silicate} < 1$. These values are actually not particular high. \citet{Hao07} and \citet{Hon10a} showed that most Seyfert 2 AGN cover a range from about $0 < \tau_\mathrm{silicate} < 1$. The Seyfert 2 mean (median) is $\left<\tau_\mathrm{silicate}\right> = 0.6$ (0.2) \citep[][note their reverse definition of emission and absorption]{Hao07} which is exactly the same as for the low-scattering sources in this paper. Deeper silicate absorption features are generally associated with cold dust in the host galaxy \citep[e.g.][]{Gou12} or, at least, at large distances from the AGN as in ULIRGs \citep{Hao07}; see also the discussion in Sect.~\ref{sec:eso103}.

Possible evidence for a deeply buried AGN could come from the mid-IR spectral index. If the AGN is deeply buried in optically very thick dust, the emerging IR emission should be redder than for normal Seyferts, since the effective temperature at which the radiation escapes is cooler. \citet{Hao07} used Spitzer spectra and determined ratios between the 5.5\,$\micron$ and 14.5\,$\micron$ fluxes for Seyfert 2 nuclei that correspond to a spectral index of approximately $-2.5 < \alpha < -0.9$, very similar to the range covered by their Seyfert 1s. Based on ground-based high angular resolution mid-IR spectra of a sample of nearby Seyfert galaxies, we found a range for mid-IR spectral indices of $-3.0 < \alpha_\mathrm{MIR} < -1.4$ with none of the objects being classified as a low-scattering source \citep{Hon10a}. As shown in Table~\ref{tab:sample}, the (dereddened) mid-IR spectral index of the decomposed AGN emission of the low-scattering objects covers about $-2.8 < \alpha^\mathrm{dereddened}_\mathrm{MIR} < -1.0$, which is consistent with the range of normal Seyferts.

We conclude that the observed $10\,\micron$ silicate absorption features of low-scattering AGN cover about the same values as those of normal Seyfert 2 galaxies. The largest $\tau_\mathrm{silicate}$ is consistent with objects where significant part of the absorption occurs in the host galaxy. Moreover we do not find any evidence for a low effective temperature of the mid-IR emission of low-scattering objects. Such a low temperature with the associated red mid-IR spectral index could be expected if an object is deeply buried within a geometrically and optically very thick shell. However, it is not found here.

\section{The low-scattering phenomenon in the context of host-galaxy absorption} \label{sec:newmod}
 
Our analysis strongly implies that low-scattering AGN have the same IR properties as other Seyfert 2 galaxies and that the only notable difference is the preferred high inclination of the host galaxies. This suggests that absorption in the host galaxy plays an important role in shaping the appearance of an obscured AGN as a low-scattering object in the present sample. Based on the finding in Sect.~\ref{sec:discu}, we want to put forward the idea to the current discussion that cold, neutral dust and gas in the host galaxy of low-scattering AGN absorbs and/or scatters the soft part of the AGN X-ray emission that arises from scattering, photoionization and/or thermal radiation. 

\begin{figure}
\begin{center}
\includegraphics[width=0.39\textwidth,angle=-90.]{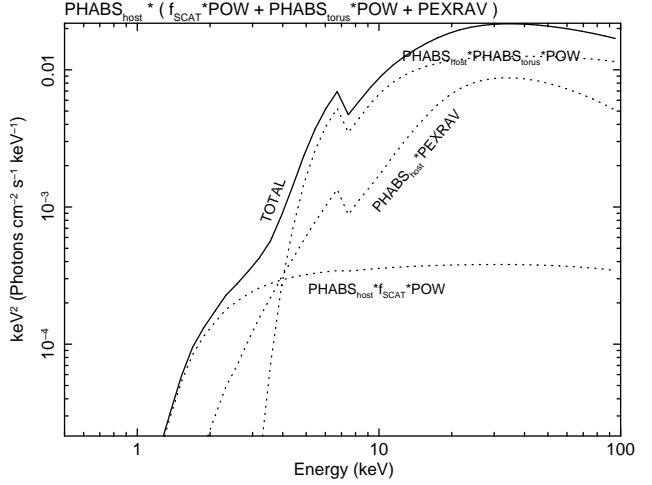}
\end{center}
\caption{XSPEC model of a host-obscured AGN that mimics the perceived low soft X-ray scattering fraction. As in normal type 2 AGN, the intrinsic AGN emission X-ray power-law emission (POW; photon index $\Gamma = 1.9$; high-energy cutoff at 300\,keV) is absorbed by the torus (PHABS$_\mathrm{torus}$); $N_\mathrm{H} = 5\times10^{23}\,\mathrm{cm}^{-2}$). In addition, the AGN environment produces a reflected component (PEXRAV; 2$\pi$ reflection) and an intrinsic scattered component ($f_\mathrm{SCAT}$*POW) with a scattering fraction $f_\mathrm{SCAT}=3\%$. Then, this total AGN spectrum ($f_\mathrm{SCAT}$*POW + PHABS$_\mathrm{torus}$*POW + PEXRAV) passes through host-galactic gas and dust (PHABS$_\mathrm{host}$) with $N_\mathrm{H} = 2\times10^{22}\,\mathrm{cm}^{-2}$.}\label{fig:xabs}
\end{figure}
 
Based on the physical parameters determined for ESO~103--G35, we estimate that the host absorption component could have a Hydrogen column density of the order of $N_\mathrm{H}\sim 10^{22}\,\mathrm{cm}^{-2}$. Without the intention to do detailed modeling of specific objects, we simulated an XSPEC spectrum\footnote{The capital variables in brackets in the following description provide the parameters set in XSPEC we used to obtain the model spectrum. They can be used to reproduce our result and start parameter studies.} of a host-obscured type 2 AGN ($\log L_\mathrm{2-10\,\keV} = 43.4$ at redshift $z=0.02$) that mimics the perceived low soft X-ray scattering fraction. This spectrum is shown in Fig~\ref{fig:xabs}. As in normal type 2 AGN, the intrinsic AGN X-ray power-law emission (photon index $\Gamma = 1.9$; high-energy cutoff at 300\,keV) is absorbed by the torus (PHABS$_\mathrm{torus}$) with a column density $N_\mathrm{H} = 5\times10^{23}\,\mathrm{cm}^{-2}$ (as typically seen in low-scattering objects). In addition, the AGN environment produces a reflected component (PEXRAV; 2$\pi$ reflection) and an intrinsic scattered component ($f_\mathrm{SCAT}$*POW) with a scattering fraction $f_\mathrm{SCAT}=3\%$, similar to what has been found in many Seyfert 1 and 2 galaxies \citep[e.g.][]{Tur97,Gua05,Cap06}. Then, this total AGN spectrum ($f_\mathrm{SCAT}$*POW + PHABS$_\mathrm{torus}$*POW + PEXRAV) passes through host-galactic gas and dust (PHABS$_\mathrm{host}$) with $N_\mathrm{H} = 2\times10^{22}\,\mathrm{cm}^{-2}$. The resulting observed soft X-ray flux would be $F_\mathrm{obs}(0.5-2\,\keV) = 6.0 \times 10^{-14}\,\ergs/\mathrm{cm}^2$ while the source emits $F_\mathrm{int}(0.5-2\,\keV) = 2.2 \times 10^{-11}\,\ergs/\mathrm{cm}^2$ intrinsically. This leads to an effective/observed scattering fraction $f_\mathrm{scat}(\mathrm{obs}) = 0.27\%$, which is consistent with classification as a low-scattering AGN.

The flux levels in the $0.5-2\,\keV$ region may be further reduced by a factor of $>$10 when increasing the host absorption by a factor of 3. In that case any additional, much weaker scattering component from the very largest scales, or any AGN photoionization component or thermal emission from the host, will be able to stand out below 3\,keV and mimic the spectra shown in \citet{Egu09}. We note again that this particular model is for illustration purposes only. Actual tests will require modeling of the detailed X-ray emission from a whole sample of low-scattering AGN (or an average spectrum) and thorough exploration of the parameter space. In addition, better spectral constraints below 1\,keV are needed for the low-scattering objects.

Host galaxy interference with the line-of-sight to the nucleus could also explain the results by \citet{Ich12}. As mentioned before, these authors found that low-scattering AGN potentially show increased star-formation in their (nuclear) mid-IR emission in WISE as compared to other type 2 AGN. As previously mentioned, the host galaxies of type 2 AGN do not show any preferred inclination of their host galaxies \citep{Kin00,Win09}. With or without circumnuclear star-formation actually present, a small subset of these type 2 hosts will be seen edge-on so that host star-formation may be projected onto the nucleus. This will show up as a weak feature of star-formation in composite spectra \citep[e.g.][]{Hao07}. Low-scattering AGN are a subset of type 2 AGN being preferentially hosted in edge-on galaxies (or mergers). In these systems the probability of having host star-formation projected onto the nucleus is much higher than for the total type 2 class that includes all inclinations (i.e. also including low inclinations where the host is not projected onto the nuclear region). Therefore, the potentially increased star-formation in low-scattering objects may just be a selection effect. We note that the reverse argument may be true for the perceived lower star-formation rate in type 1s as compared to type 2 AGN: According to \citet{Kee80}, \citet{Kin00} and \citet{Win09}, AGN classified as type 1 have a deficiency of highly-inclined host galaxies which would reduce the likelihood of projecting host star-formation onto the nucleus. Both examples illustrate the importance of host galaxy absorption/projection and related selection effects in AGN classification.

An open issue  is the presumed low [\ion{O}{iii}] equivalent width in low-scattering AGN. Based on our host-absorption scenario, we would expect significant extinction toward the NLR \citep[e.g. as reported by][]{Kir90,Haa05,Tov11,Kra11}. We encourage follow-up studies along this line. Of particular interest is a comparison of optical emission lines to high-ionization mid-IR AGN tracer lines, which could reveal potential host absorption toward the NLR. It is worth noting that for ESO~005--G4 \citet{Wea10} found [\ion{Ne}{v}](14.32\,$\micron$) and [\ion{O}{iv}](25.89\,$\micron$) luminosities that are typical for other Seyfert galaxies. They conclude that the missing optical high-ionization lines are probably caused by host-galactic dust since this pattern is also sometimes found in other AGN hosted by highly-inclined galaxies that are not classified as low-scattering. If this holds true for the other low-scattering objects, it would be a strong indication that dust and gas in the host galaxy is indeed responsible for the appearance of an AGN as a low-scattering object. The host Hydrogen column densities of the order of $N_\mathrm{H}\sim10^{22}\,\mathrm{cm}^{-2}$ required to produce the soft X-ray-deficient spectra would provide ample obscuration in the optical. For typical ISM gas and dust composition, a conversion of column density to dust optical depth is found as $\tau_V \sim 2 \times N_\mathrm{H}/10^{21}\,\mathrm{cm}^{-2}$ \citep[e.g.][]{Fit85}, meaning we could expect $\tau_V$ of the order of several 10s. If the emission from the NLR is predominantly originating from its innermost regions \citep[few parsec;][]{Pet13}, these values are probably sufficient to extinct most of the high-ionization lines. In fact, the expected $\tau_V$ corresponds to a silicate optical depth $\tau_\mathrm{silicate}$ of order unity and, therefore, being in line with the observed absorption features that are interpreted as arising from cool dust \citep[see also discussion in][]{Gou12}.

\section{Summary and conclusions}\label{sec:summary}

Our goal was to study AGN that have been identified in X-ray surveys for their very low observed soft-X-ray scattering fraction $f_\mathrm{scat} < 0.5$\%. Based on X-ray modeling, these ``X-ray new-type'', or \textbf{``low-scattering''} objects are supposedly buried in a dusty torus that has a very small opening angle. This would result in a small solid angle available for X-rays escaping from the AGN environment. In particular,  we addressed if the galaxies that host these sources show any preferred characteristics, and if we can find any evidence for this scenario in the IR radiation of these sources.

The main conclusions from our study are:
\begin{itemize}
\item We find evidence that low-scattering AGN are preferentially hosted in edge-on galaxies or mergers. Most strikingly, while for a random host orientation we would expect 17\% of galaxies having inclination $i>80^\circ$, the low-scattering sample show these high inclinations at a frequency of 50\%. In such systems, it is very likely that dust and gas in the host galaxy intersects with our line-of-sight to the nucleus.
\item By comparing the X-ray and IR emission, we did not find any evidence that the dust covering factors in the low-scattering AGN in our sample are any different than in Seyfert nuclei that have the same X-ray column density. 
\item We also found that the observed $10\,\micron$ silicate absorption features and mid-IR spectral indices of low-scattering objects cover about the same range as other type 2 AGN.
\item For one low-scattering object, ESO~103--G35, we were able to determine several physical properties of the dust. We found a (bolometric) covering factor $C_\mathrm{bol} = 0.5^{+0.5}_{-0.3}$. While this direct determination of a covering factor based on incident and reemitted radiation could be a useful test for the low-scattering picture, the errors do not allow to put tight constraints on the covering factor in this case. We conclude, however, that the IR emission originates from warm dust with a mass of $M_\mathrm{dust} = (1.4\pm0.4) \times 10^4\,\mathrm{M_\odot}$ at a distance of about 13\,pc from the AGN, consistent with a dusty torus or other circumnuclear dust distributions discussed in literature. The observed silicate absorption feature suggests additional absorption by cold dust (temperature $\la 50$\,K) in the host galaxy, constrained to have $N_\mathrm{H}$ of the order of $10^{22}$\,cm$^{-2}$ and being consistent with the high inclination of the host galaxy.
\end{itemize}

In summary, our results imply that in the present sample the high inclinations of the host galaxies is the major difference between low-scattering AGN and other Seyfert galaxies that cover the same parameter space of AGN luminosity, line-of-sight column densities, or IR reemission. This suggests that absorption in the host galaxy plays an important role in shaping the appearance of an obscured AGN as a low-scattering object. We present an X-ray model that qualitatively reproduces the low perceived X-ray scattering $f_\mathrm{scat} < 0.5$\%, invoking standard X-ray emission properties of a type 2 AGN plus host galaxy absorption. We strongly encourage to test this scenario in future modeling of larger samples of low-scattering objects, e.g. using the BAT 70 month sample \citep{Bau13} once spectral modeling results are available.

We have not completely ruled out the hypothesis that individual low-scattering AGN have unusually high covering factors. Support for such a scenario may arise if, in addition to the peculiar X-ray spectrum, the AGN is hosted in an early-type galaxy that is generically devoid of dust \citep[e.g.][]{Sob12}. In such cases, it would be more difficult to imagine host dust to contribute toward absorption significantly. However, the objects we investigated in this study are radio-quiet Seyfert galaxies mostly in spiral/disk galaxies or mergers that do contain large amounts of dust. Therefore we showed that a low X-ray scattering fraction is not a sufficient condition to conclude that these objects have generically high covering factors.

\section*{Acknowledgments}

We want to thank M. Melendez and T. Shimizu for providing the PACS and SPIRE photometry, respectively, of ESO~103--G35, and the anonymous referee for comments and suggestions. This work is based in part on observations made with Herschel, a European Space Agency Cornerstone Mission with significant participation by NASA. Support for this work was provided by NASA through an award issued by JPL/Caltech. The Dark Cosmology Centre is funded by The Danish National Research Foundation. P.G. acknowledges support from STFC (grant reference ST/J003697/1). This research has made use of the NASA/IPAC Extragalactic Database (NED) which is operated by the Jet Propulsion Laboratory, California Institute of Technology, under contract with the National Aeronautics and Space Administration. We acknowledge the usage of the HyperLeda database (http://leda.univ-lyon1.fr). This publication makes use of data products from the Two Micron All Sky Survey, which is a joint project of the University of Massachusetts and the Infrared Processing and Analysis Center/California Institute of Technology, funded by the National Aeronautics and Space Administration and the National Science Foundation. This publication makes use of data products from the Wide-field Infrared Survey Explorer, which is a joint project of the University of California, Los Angeles, and the Jet Propulsion Laboratory/California Institute of Technology, funded by the National Aeronautics and Space Administration.

\footnotesize{
  \bibliographystyle{mn2e}

}

\bsp

\label{lastpage}

\end{document}